\documentclass[twocolumn,eqsecnum,aps,prd,showpacs,preprintnumbers,amsmath,amssymb,nofootinbib,floatfix, superscriptaddress]{revtex4}

\usepackage[section]{placeins}
\usepackage{bm}
\usepackage{graphicx,epsf}
\usepackage{pstricks}
\usepackage{amssymb}

\newcommand\calp{\mathcal{P}}

\newcommand\mpl{m_{\rm Pl}}

\renewcommand\({\left(}
\renewcommand\){\right)}
\newcommand\be{\begin{equation}}
\newcommand\ee{\end{equation}}
\newcommand\bea{\begin{eqnarray}}
\newcommand\eea{\end{eqnarray}}
\newcommand\eq[1]{equation ~(\ref{#1})}
\newcommand\eqs[2]{equations ~(\ref{#1}) and (\ref{#2})}
\newcommand\fig[1]{Fig.~\ref{#1}}

\newcommand\eps{\epsilon}

\begin{document}
\title{Generating Primordial Black Holes Via Hilltop-Type Inflation Models}
\author{Laila Alabidi}
\email{l.alabidi@qmul.ac.uk}
\affiliation{Astronomy Unit, School of Mathematical Sciences, 
Queen Mary University of London, Mile End Road, London, E1 4NS, 
United Kingdom} 
\author{Kazunori Kohri}
\email{k.kohri@lancaster.ac.uk}
\affiliation{Department of Physics, University of Lancaster, Lancaster, LA1 4YB, United Kingdom}

\begin{abstract}
	It has been shown that black holes would have formed in the early Universe
	if, on any given scale, the spectral amplitude of the Cosmic Microwave Background (CMB)
	exceeds $\calp_{\zeta}\sim10^{-4}$. This value is within the bounds allowed
	by astrophysical phenomena for the small scale spectrum of the CMB, corresponding to
	scales which exit the horizon at the end of slow-roll inflation. Previous
	work by Kohri et. al. (2007) showed that for black holes to form from a single field
	model of inflation, the slope of the potential at the end of inflation must be flatter than it was
	at horizon exit. In this work we show that a phenomenological Hilltop model of inflation,
	satisfying the Kohri et. al. criteria, could lead to the production of black holes, 
	if the power of the inflaton self-interaction 
	is less than or equal to $3$,  with a reasonable number or $e-$folds. We
	extend our analysis to the running mass model, and confirm that this model 
	results in the production of black holes, and by using the latest
	WMAP year 5 bounds on the running of the spectral index, and the black hole constraint
	we update the results of Leach et. al. (2000) excluding more of parameter space. 
\end{abstract}

\pacs{}

\maketitle

\section{Introduction}

The temperature anisotropies of the Cosmic Microwave Background (CMB)
have been measured by WMAP on  angular scales down to
$\theta\sim0.3^o$, whereas they have yet  to be measured to an
effectual degree of accuracy on scales $\theta<0.3^o$
\cite{Nolta:2008ih}.  In fact, CMB data  naively allows for a very
large spectrum on these scales, i.e. a spectrum a few orders of
magnitude above
$\calp_{\zeta_*}\simeq10^{-9}$, evaluated at horizon exit. One can, however, place an upper limit
on the smaller scale spectrum by taking into account astrophysical and
cosmological constraints on black holes
\cite{Carr:1994ar,Kim:1996hr,Green:1997sz,Green:1999yh,
Kohri:1999ex,Lemoine:2000sq,Barrau:2001ev,Barrau:2002ru,Josan:2009qn,Carr:2009}.  If this spectrum  is in
fact close to this upper limit, then the situation allows for `large'
fluctuations; large enough to collapse into black holes, known as
Primordial Black Holes (PBHs)
(c.f. \cite{1967SvA....10..602Z,Hawking:1971ei,Carr:1974nx,
Carr:1975qj,Khlopov:1985jw,Ivanov:1994pa,GarciaBellido:1996qt}).
Since the uncertainties in the primordial spectrum at such small
scales are dominated by instrumental noise \cite{Nolta:2008ih},  as
opposed to cosmic variance which is the dominant source of
uncertainties at larger angular scales, it may be that these
uncertainties can be reduced in future surveys. Therefore, the
question of whether the spectrum of perturbations on small scales is
large enough to form PBHs is one that can, in theory, be
answered.

The $\theta\gtrsim0.3^o$ spectrum has been used extensively in
discriminating between models of inflation
(c.f. \cite{Kinney:2003uw,Alabidi:2005qi,Alabidi:2006qa,Kinney:2006qm,Martin:2006rs,Kinney:2008wy,Alabidi:2008ej}).
These analyses are based on the assumption that the anisotropies in
the CMB, and hence the origin of large scale structure, are sourced by
quantum fluctuations in a scalar field during or straight after
inflation. As such, we assume that the fluctuations which sourced the
PBHs are also generated during  inflation, specifically towards the
end of inflation, as it is during this epoch that the small scale
fluctuations exit the horizon.  Measuring the $\theta<0.3^o$ spectrum
will therefore not only probe generic signatures of inflation
\cite{Carr:1994ar,Lidsey:1995ir} but also act as an indicator for the
shape of the inflationary potential
\cite{GarciaBellido:1996qt,Kohri:2007qn}. In this paper we aim to
exploit the latter purpose of measuring the small scale spectrum,  and
investigate whether single field models  of inflation can lead to the
formation of PBHs.~\footnote{For multi-field or multi-stage
inflation models to produce PBHs, see
Refs.~\cite{Yokoyama:1995ex,Kawasaki:1997ju,
Yokoyama:1998pt,Yokoyama:1998qw,
Kawasaki:1998vx,Yokoyama:1999xi,Kanazawa:2000ea,Yamaguchi:2001zh,
Blais:2002nd,Kawasaki:2007zz,Kawaguchi:2007fz,Saito:2008em} and
references therein.}

To be specific, the primordial black hole (PBH) condition 
\cite{Carr:1974nx,Kohri:1999ex,Carr:1975qj,Carr:2005bd,Zaballa:2006kh}
can be expressed as:
\be\label{pbh_spectrum}
\calp_{\zeta_e}^{1/2}\simeq0.03=10^3\calp_{\zeta_*}^{1/2}
\ee
where the subscripts $e$ and $*$ refer to the end
of inflation and horizon exit respectively. Assuming a constant spectral index then:
\be\label{n_constant}
n_s-1\simeq\frac{d\ln\calp_\zeta}{dN}\simeq\frac{\Delta\ln\calp_\zeta}{\Delta{}N}\simeq\frac{14}{\Delta{}N}.
\ee
where we used (\ref{pbh_spectrum}) in the last semi-equality.
$\Delta{}N\sim{}N$ refers to the number of $e-$folds that
elapse from the time when scales of cosmological interest
leave the horizon till the end of inflation.

Taking a standard value of $N\simeq60$, it then
follows from (\ref{pbh_spectrum})
that $n_s\simeq1.3$. This is beyond the 
upper limit of $0.993$ allowed by the recent WMAP data,
at $95\%$ confidence  limit
with no tensor modes or running. Therefore we consider 
the variation in the spectrum up to second order, and 
assume that the spectral index depends on scale, 
i.e. $\calp\propto{}k^{n_s(k)-1}$, then:

\bea
\frac{d\ln{}\calp}{d\ln{}k}&=&(n_s-1)+n_s'\ln{}k\nonumber\\
\ln\left[\frac{\calp_{\zeta_e}}{\calp_\zeta(N)}\right]&=&N(n_s-1)+\frac{1}{2}n_s'N^2\\
&=&14
\eea
where we used $\ln{}k=\ln(aH)=N$ in the second step. 
Taking $n_s\simeq0.95$ and $N=60$ then requires
$n_s'\simeq0.0061$ to satisfy the WMAP bounds 
and produce Primordial Black Holes.

We now wish to rewrite the Primordial Black Hole Condition
(\ref{pbh_spectrum}) in terms of the slow roll parameters.
Recalling that the spectrum can be written in terms of 
the potential $V$ and the slow roll parameter $\epsilon=\mpl^2(V'/V)^2/2$, then \cite{Liddle:2000cg}:
\be\label{spectrum_epsilon}
\calp_\zeta=\frac{1}{24\pi^2\mpl^4}\frac{V}{\epsilon}
\ee
Defining a new parameter $\mathcal{B}=\epsilon_e/\epsilon_*$, and combining
\eqs{pbh_spectrum}{spectrum_epsilon} we find that the condition for PBH formation,
without violating the aforementioned astrophysical and cosmological bounds is:
\be\label{pbh_cond}
\mathcal{B}\simeq10^{-6}
\ee
An `absolute' upper bound on the spectrum is given by \cite{Josan:2009qn} as $\calp_{\zeta_e}^{1/2}\sim10^{-1}$,
which translates to a lower bound $\mathcal{B}>10^{-8}$.

Equation (\ref{pbh_cond}) tells us that for an inflationary potential to
lead to the production of PBHs, its' slope must flatten towards the end of inflation.
This shape is satisfied by
a phenomenological model akin to the one analysed in
\cite{Kohri:2007gq}, and also the running mass model, first introduced
in \cite{Stewart:1996ey}.  In these types of models, we require that the
inflaton initially be sitting
at around the top of the hill, that is near a local maxima. 
This condition can be considered natural \cite{Kohri:2007gq} from the viewpoint 
of eternal inflation \cite{Vilenkin:1983xq}, and can be understood as follows: via some mechanism,
be it quantum tunnelling or an inhomegenous pre-inflationary
universe, the inflaton will somewhere, at some time find itself
sitting at the top of the potential, at which point the universe
will start to inflate. As long as the inflaton is undisturbed, the universe
will inflate indefinatley, and can end up volumetrically dominating the universe.
Since this process can lead to an indefinatley large volume, then even
if there was the smallest probability that inflation were to start, it would \cite{Guth:2007ng, Lyth:2009book}.
Within this patch, quantum fluctuations in some sub-patches displace the inflaton 
from its vestige causing
it to roll either to the left or the right, and ending inflation in those regions, while
overall, the patch continues to inflate \cite{Guth:1985ya, Guth:2007ng}. 
We introduce these models in sections
(\ref{sec_tree}) and (\ref{sec_run}) respectively. We present our
results in section (\ref{sec_result}) and discuss them in section
(\ref{sec_discuss}).

\subsection{The Number of $e-$folds}\label{N_sec}

In this paper we use the duration of inflation,
otherwise known as the number of $e-$folds $N$,
as a discriminator. It is defined as the ratio
of the scale factor $a$ at the end of inflation
to $a$ at the `beginning' of inflation:
\be
\label{N_eq}
N=\ln\left[\frac{a_e}{a_*}\right]\simeq\mpl^{-1}\int_{\phi_e}^{\phi_*}\frac{d\phi}{\sqrt{2\epsilon}}
\ee
where the final semiequality comes from the slow
roll approximation.

To get a proper handle on how long inflation
lasted from the time of horizon exit, one needs
a complete history of the Universe. At present
though, we do not have an agreed upon mechanism of reheating. 
Therefore, one assumes an instant transition from inflation
to a radiation dominated universe, and gets
the bounds \cite{Liddle:2003as}:

\be
\label{N_limit1}
10\lesssim{}N\lesssim110
\ee
The lower bound comes from
the assumption that Nucleosynthesis is well bounded, 
and the upper bound assumes
that the universe underwent a few bouts of `fast' roll
inflation. We do note that these are extreme bounds,
and that the limits $N=54\pm7$ are more widely acceptable 
(c.f. \cite{Liddle:2000cg, Liddle:2003as, Alabidi:2005qi}).

\subsection{Slow Roll and Cosmological Parameters}
The lowest order slow order parameters are given by:

\bea\label{SR}
\eps&=&\frac{\mpl^2}{2}\left(\frac{V'}{V}\right)^2\\
\eta&=&\mpl^2\frac{V''}{V}\\
\xi^2&=&\mpl^4\frac{V'V'''}{V^2}\\
\eea
which we then use to compute the spectral index and
the running:
\bea
n_s&=&1+2\eta-6\eps\\
n_s'=\frac{dn}{d\ln{}k}&=&16\eps\eta-24\eps^2-2\xi^2
\eea

We use the bounds on cosmological parameters 
given by the WMAP year 5, Baryon Acoustic Oscillations and 
Supernovae data sets  \cite{Komatsu:2008hk}:

\bea\label{cosmobounds}
0.939<n_s<1.109\\
-0.0728<n_s'<.0087
\eea
where a zero tensor mode was assumed in the prior. 
This prior is reasonable since we are considering small field models,
characterised by a field variation that is smaller than
the Planck mass $\Delta\phi<\mpl$. In these models the gravitational wave
signature will be small \cite{Lyth:1996im}, and by small we mean well below the
sensitivity of WMAP5 parameter estimation, so we do not calculate
the associated parameter.

\section{The Tree Level Potential}\label{sec_tree}

\begin{figure*}
\includegraphics[width=\linewidth, totalheight=4in]{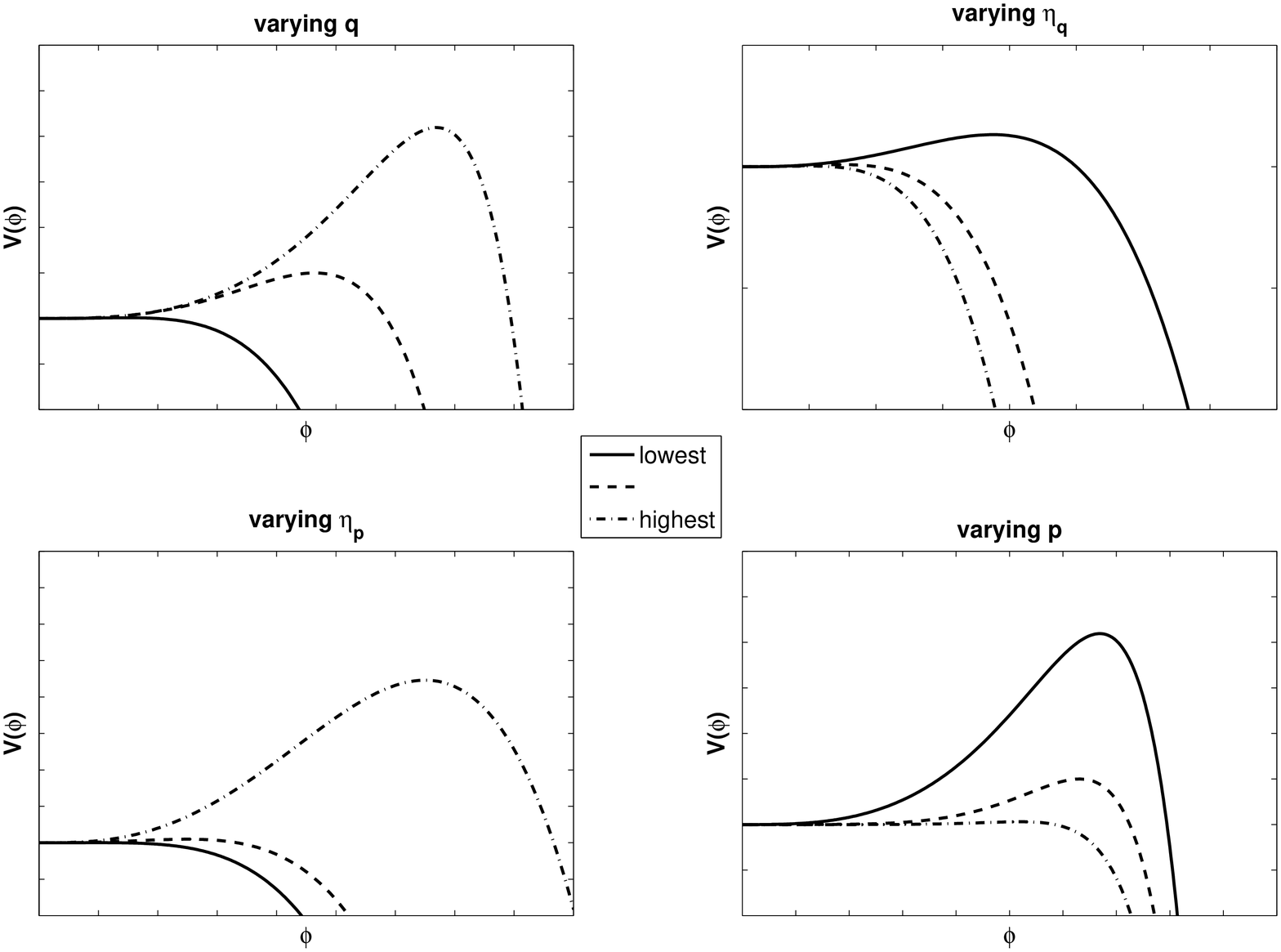}
\caption{For the figures on the left, increasing the variant decreases $N$, and the opposite is true for the variants defining the figures on the right.}
\label{potential_figure}
\end{figure*}

We consider the potential of the form:
\be\label{potential}
V=V_0\left(1+\eta_p\(\frac{\phi}{\mpl}\)^p-\eta_q\(\frac{\phi}{\mpl}\)^q\right)
\ee
where $p\geq2$ and $q>p$, plotted in \fig{potential_figure}. 
The case $p=2$ and $q\geq4$ can be generated
from a flat direction in the Minimal Supersymmetric Model (MSSM) 
\cite{Allahverdi:2006iq,Lyth:2006ec,Allahverdi:2006we,BuenoSanchez:2006xk,Lin:2009yt,Lin:2009ux}. 
In this case $q$ is the non-renormalisable operator that depends 
on the flat directions. One can also motivate the parameter range $p=2$
and $q\geq3$ in \cite{Allahverdi:2006cx,Lin:2009ux,Kohri:2009ka}, in this scenario
the inflaton higher order terms are not Planck suppressed, and one gets
a lower energy scale inflation, on the order of the TeV -- GUT scale.
We also maintain $\phi<\mpl$, a realistic bound from an effective
particle physics perspective, which demands that one not consider mass scales
larger than the largest naturally occurring scale, in this case the Planck mass. 
Then due to the Lyth bound, the gravitational wave contribution of this model is negligible,
regardless how long inflation lasts.

In this setup we require that the potential at the end of 
inflation be flatter than it was at the time of horizon exit, 
so the inflaton must roll towards the origin. We denote 
the inflaton value at the maximum of the potential as 
$\phi_m$, at horizon exit as $\phi_*$ and at the end of 
inflation as $\phi_e$. We impose the conditions:

\bea\label{limits}
        \phi_*<\phi_m\\
        0<\phi_e<\phi_o
 \eea
where $\phi_o$ is the inflection point. $\phi_m$ and $\phi_o$ are then given by (\ref{potential}):

\bea
\frac{\phi_m}{\mpl}&=&\left(\frac{p\eta_p}{q\eta_q}\right)^{1/(q-p)}\\
\phi_o&=&\phi_m\left(\frac{p-1}{q-1}\right)^{1/(q-p)}
\eea

\section{The Running Mass Model}\label{sec_run}

With the exception of $p=2$ and integral values $3\leq{}q\leq9$, 
the previous model is a phenomenological one.
However the shape of potential does appear 
in a more theoretically motivated setup, the running mass model \cite{Stewart:1996ey,Stewart:1997wg,Covi:1998jp,Covi:1998mb,Covi:2000qx,Lyth:2000qp,Covi:2002th,Covi:2004tp,Leach:2000ea}
which has the potential:
\be
V=V_0\left[1-\frac{1}{2}\mu^2\frac{\phi^2}{\mpl^2}\right]
\ee
where the mass of the inflaton is scale dependent and can be expressed as:
\be\label{running_mass}
\mu^2(\phi)=\mu_0^2+A_0\left[1-\frac{1}{(1+\alpha\ln{(\phi/\mpl)})^2}\right]
\ee
where $\mu_0^2$ is the mass of the
inflaton squared, $A_0$ is the gaugino mass squared in units of $\mpl$, and $\alpha$ is related
to the gauge coupling. 

\begin{figure}[ht]
\centering
\includegraphics[width=\linewidth]{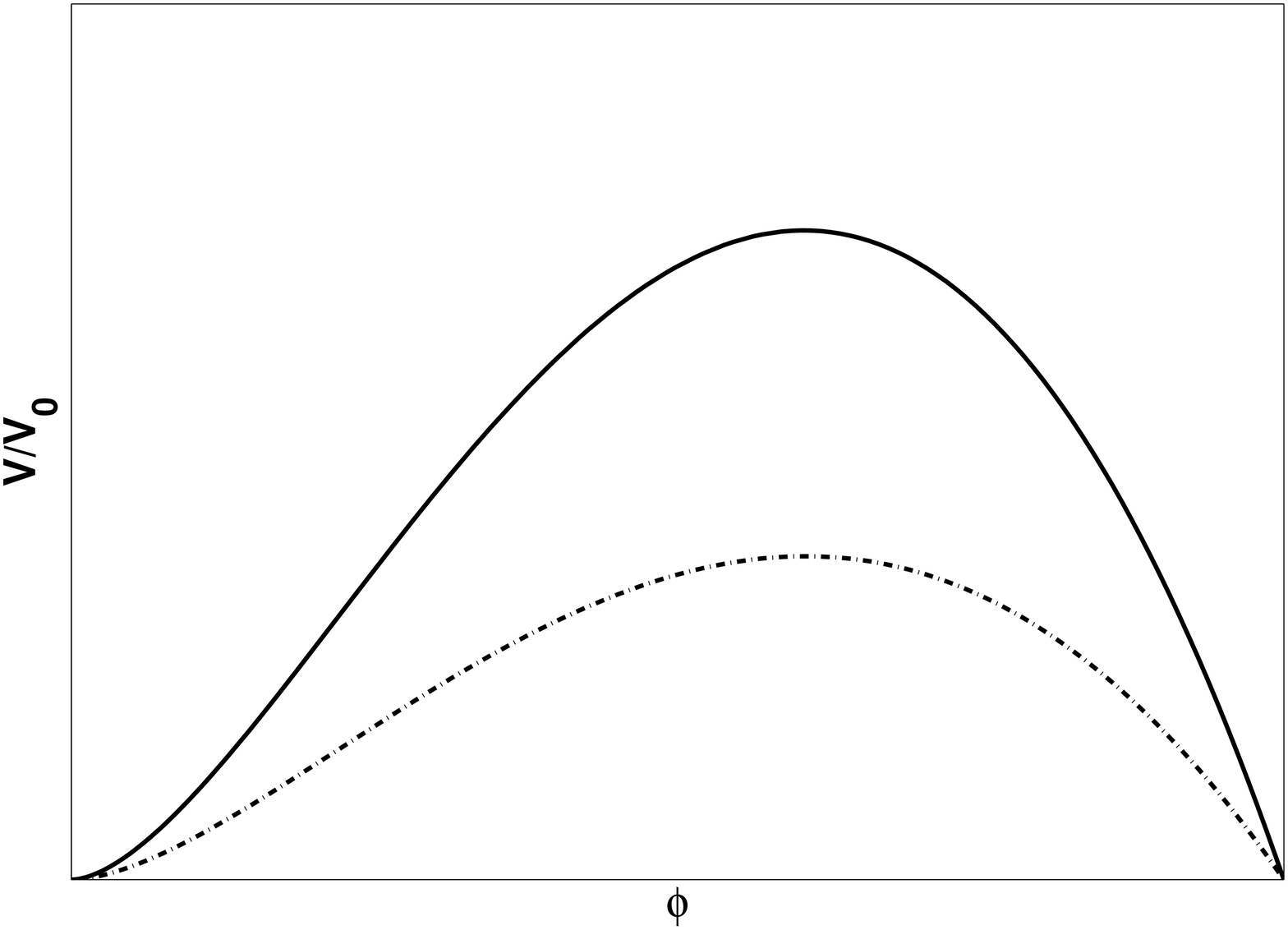}
\caption{Plot of the running mass model. The solid line represents a larger $\alpha$ than the dashed-dotted line.}
\label{RMM_fig}
\end{figure}

The potential can then be written as:
\be\label{RMM}
V=V_0\left[1-\frac{1}{2}B_0^2\left(\frac{\phi}{\mpl}\right)^2+\frac{A_0}{2(1+\alpha\ln(\phi/\mpl))^2}\left(\frac{\phi}{\mpl}\right)^2\right]
\ee
and $B_0^2=\mu_0^2+A_0$, with inflation occurring in the regime $\phi\ll\mpl$. 
This potential has the shape in \fig{RMM_fig}, and the parameters have
theoretically motivated constraints \cite{Covi:1998jp}:
\bea\label{RMM_constraint}
1\lesssim{}\mu_0^2\lesssim\mathcal{O}(10)\nonumber\\
0\lesssim{}A_0\lesssim\mathcal{O}(10)\nonumber\\
10^{-3}\lesssim\alpha\lesssim10^{-1}
\eea

\subsection{Linear Approximation}

The linear approximation of the running mass potential is given by \cite{Covi:1998mb,Kohri:2007qn}:

\be
\label{RMM_Covi}
\frac{V}{V_0}=1-\frac{\phi^2}{2}(\mu_{*}^{2}+c\ln(\phi/\phi_m))
\ee
which is basically \eq{RMM} expanded about the maximum of the potential. 
The terms $c$ and $\mu_*^2$ are related to the theoretical
parameters $A_0,\mu_0^2$ and $\alpha$ by:

\bea\label{para_relation}
\mu_*^2&=&\mu_0^2+A_0\left[1-\frac{1}{(1+\alpha\ln(\phi_m))^2}\right]\nonumber\\
c&=&\frac{2\alpha{}A_0}{(1+\alpha\ln(\phi_m))^3}
\eea
Since $\phi_m$ defines the maximum of the potential then $\mu_*^2=-c/2$, this
allows us to write:

\bea
N&=&-\frac{1}{c}\ln\left[\frac{\ln(\phi_m/\phi_*)}{\ln(\phi_m/\phi_e)}\right]\label{Nlinear}\\
\frac{n-1}{2}&=&-c\left[\ln\left(\frac{\phi}{\phi_m}\right)+1\right]\\
&=&\sigma{}e^{-cN}-c
\eea
where $\sigma=-c\ln(\phi_e/\phi_m)$.

This approximation is only valid near the maximum. 
We neither expect nor get reasonable values of $\phi_e$
using this estimate. However
given $\phi_*$ and $\phi_e$, we found that the linear approximation for
$N$ in (\ref{Nlinear}) appears consistent with the numerical
calculation.

\section{Results}\label{sec_result}
\subsection{Hilltop}
\begin{figure}
\centering
\includegraphics[totalheight=2.5in, width=\linewidth]{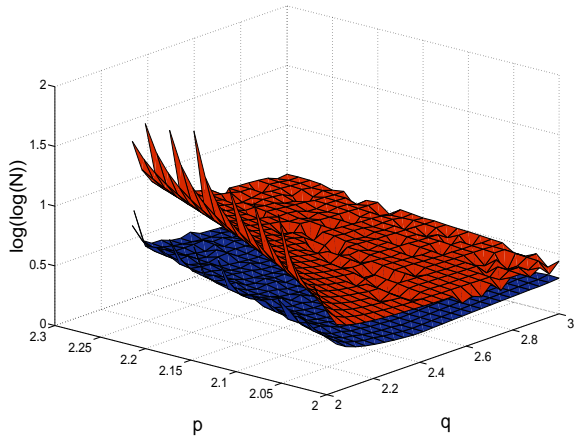}
\caption{A plot of the maximum and minimum values of $\log(\log(N))$ versus $p$ and $q$, from a range values of $\eta_p$ and $\eta_q$.}
\label{pq}
\end{figure}

For the tree level potential, \fig{pq} depicts the dependence of $N$ 
on $p$ and $q$, we note that $N$ increases
sharply as $p$ increases, as expected. We then searched for the range of $p$ and $q$ parameters
that satisfy the bounds $10<N<110$ and found that the condition 
for PBH formation within this range is $2\leq{}p<3$ and $p<q<4$. Larger values of $p$ lead
to $N\gg110$ while larger values of $q/p$ do not satisfy the WMAP
bounds, since they steepen the potential and result in an increased $n'$. We also found that 
$p\simeq2$ and $q<3$ (note $q\neq3$) places $N$ in the 
range $N=54\pm7$.

\begin{figure}
 \centering
\includegraphics[totalheight=2.5in, width=\linewidth]{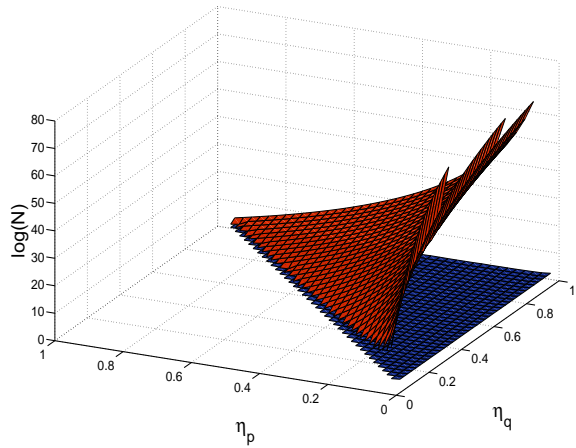}
\caption{A plot of the minimum values of $\log(N)$ versus $\eta_p$ and $\eta_q$, from a range values of $p$ and $q$.}
\label{epeq}
\end{figure}

We then plot the dependence of $N$ on $\eta_p$ and $\eta_q$ in \fig{epeq}.
As expected $N$ increases for decreasing $\eta_p$ and vice versa for $\eta_q$. Once
we have filtered out the reasonable values of $N$, we find that within
the range $\{\eta_p,\eta_q\}=\{0,1\}$ PBH formation will occur.
It seems that the stronger constraints for PBH formation
with $10<N<110$ come from $p$ and $q$.

\begin{figure*}[ht]
 \begin{minipage}{0.4\linewidth}
\centering
\includegraphics[width=\linewidth, totalheight=2in]{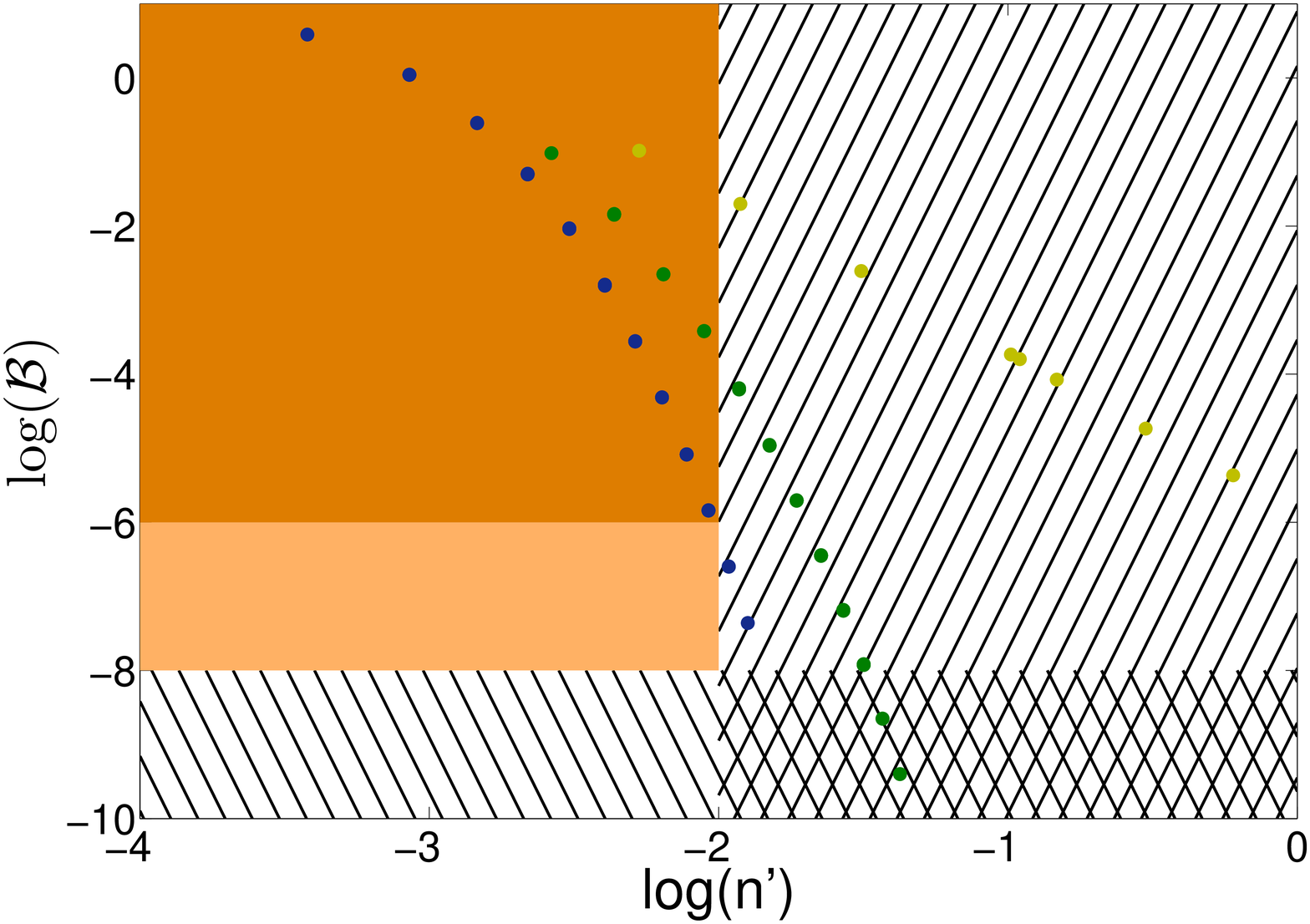}
\end{minipage}
\hspace*{0.5 cm}
\begin{minipage}{0.4\linewidth}
\centering\includegraphics[width=\linewidth,totalheight=2in]{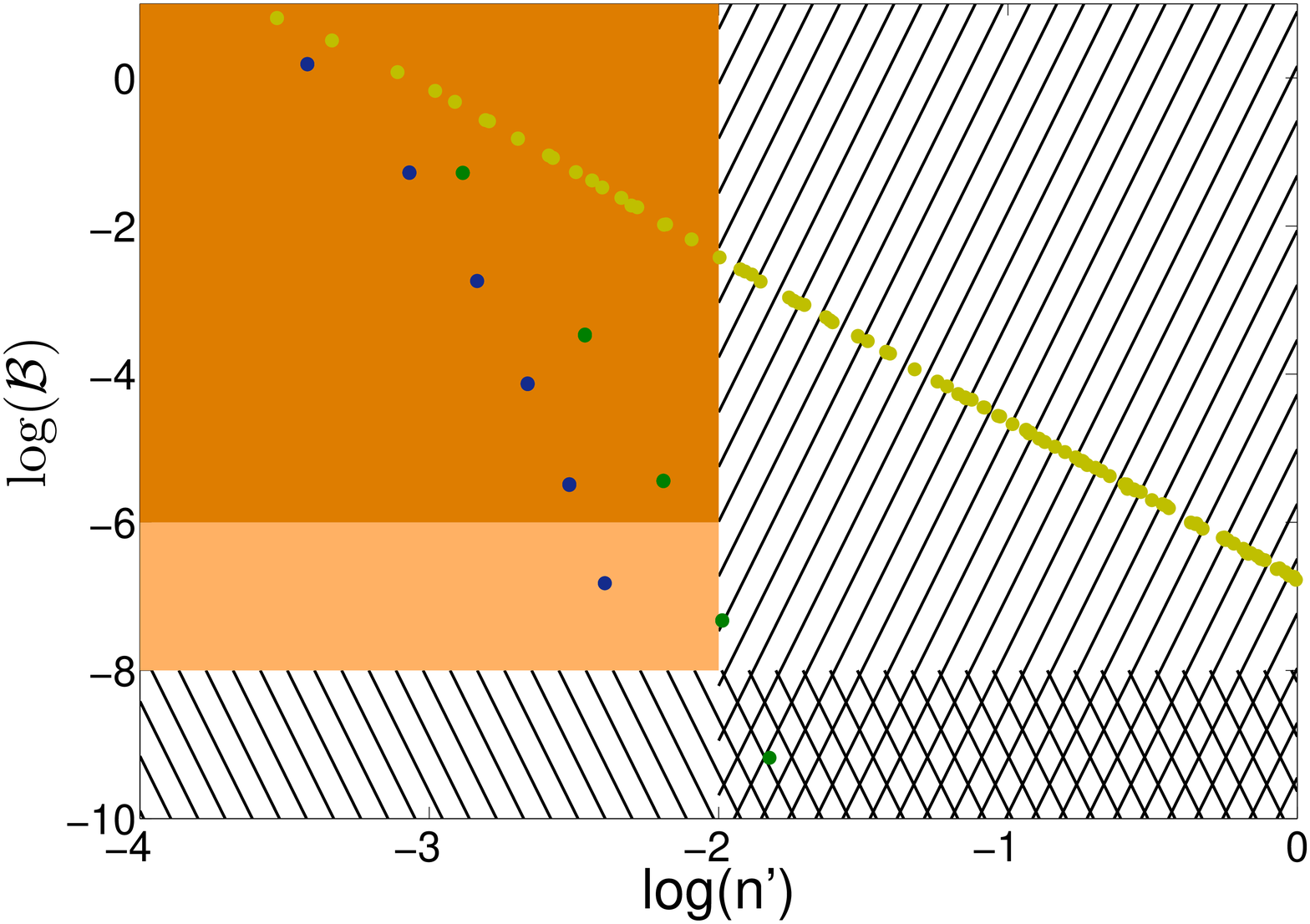}
\end{minipage}
\caption{Plot of $\log(\mathcal{B})$ versus $\log(n')$ for the Hilltop model
with $N=60$ (figure on left), $N=100$ (figure on right) and $n_s=0.95$. The hatched region is excluded, representing
$\log(n')>-2$ and $\log(\mathcal{B})<-8$. The region $\log(\mathcal{B})>-6$ does not
lead to the formation of PBHs, and is represented by the tan colour in the figure.
PBHs can form in the region $-8\leq\log(\mathcal{B})\leq-6$ without violating astrophysical
or cosmological bounds, and is represented by the light orange region. The yellow dots correspond to
$\{p,q\}=\{3,4\}$, the green dots to $\{p,q\}=\{2,3\}$ and the blue dots
to $\{p,q\}=\{2,2.5\}$.} 
\label{eta_all}
\end{figure*}

Next we consider the case of defining $\phi_*$ by the
condition that $n(\phi_*)=0.95$ and $\phi_e$ by
$N=60$ and $N=100$. \fig{eta_all} shows that for both $N=60$ and $N=100$, the parameter
ranges satisfy the WMAP bounds and the PBH constraints. However,
for $N=60$ integral values of $p$ and $q$ do not lead to the formation
of PBHs, while for $N=100$, the parameter set $\{p,q\}=\{2,3\}$ does.

\subsection{The Running Mass Model}

In this analysis we considered the parameter ranges in (\ref{RMM_constraint}),
for which we find that this model
generates a large range of $N$ values. In \fig{Nreasonable2_RMM}
we plot the allowed parameter space for $10\leq{}N\leq110$ and $N=54\pm7$.

\begin{figure*}
\begin{minipage}{0.4\linewidth}
 \centering
\includegraphics[width=\linewidth,totalheight=2in]{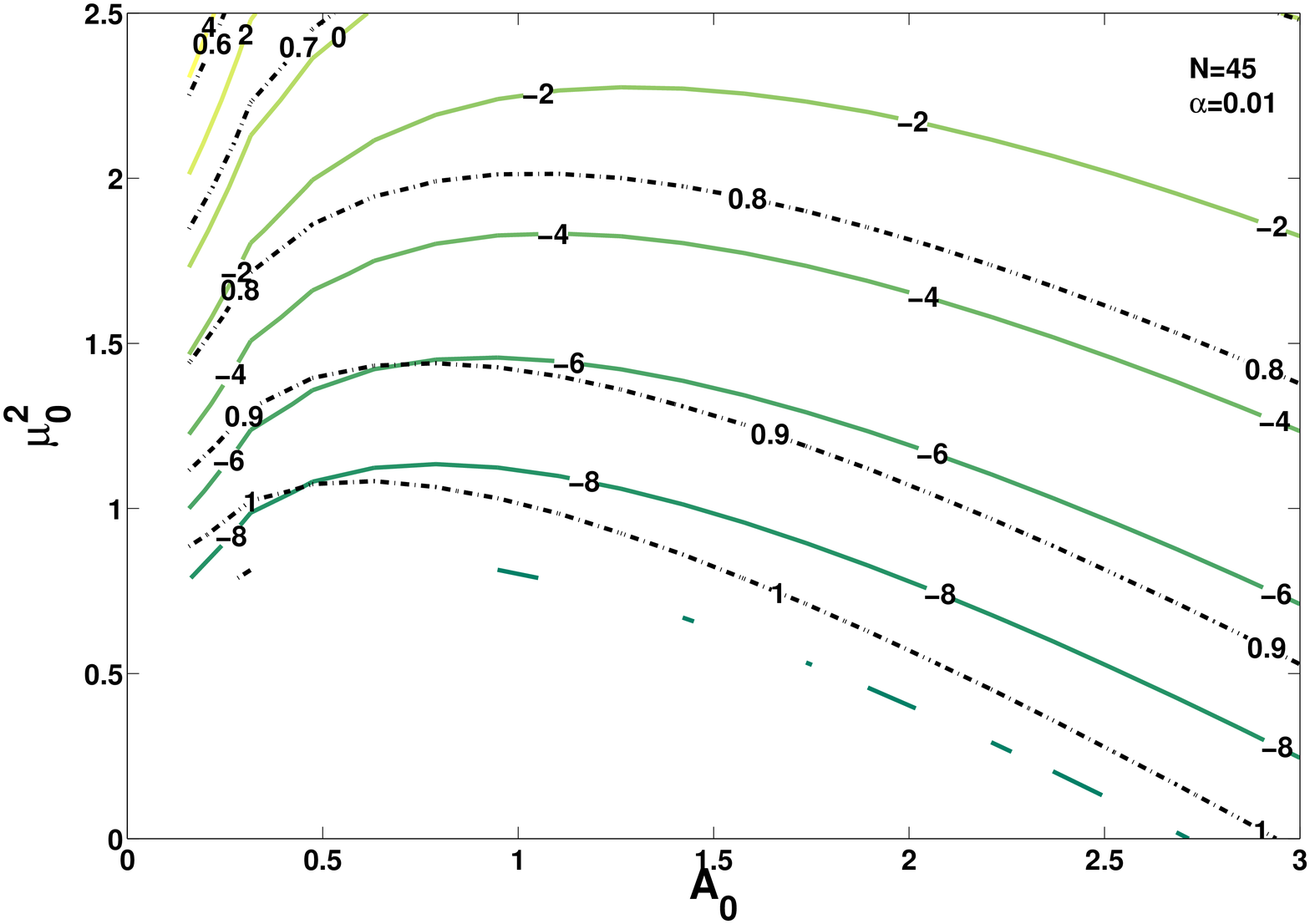}
\end{minipage}
\hspace*{0.5 cm}
\begin{minipage}{0.4\linewidth}
\centering
\includegraphics[width=\linewidth,totalheight=2in]{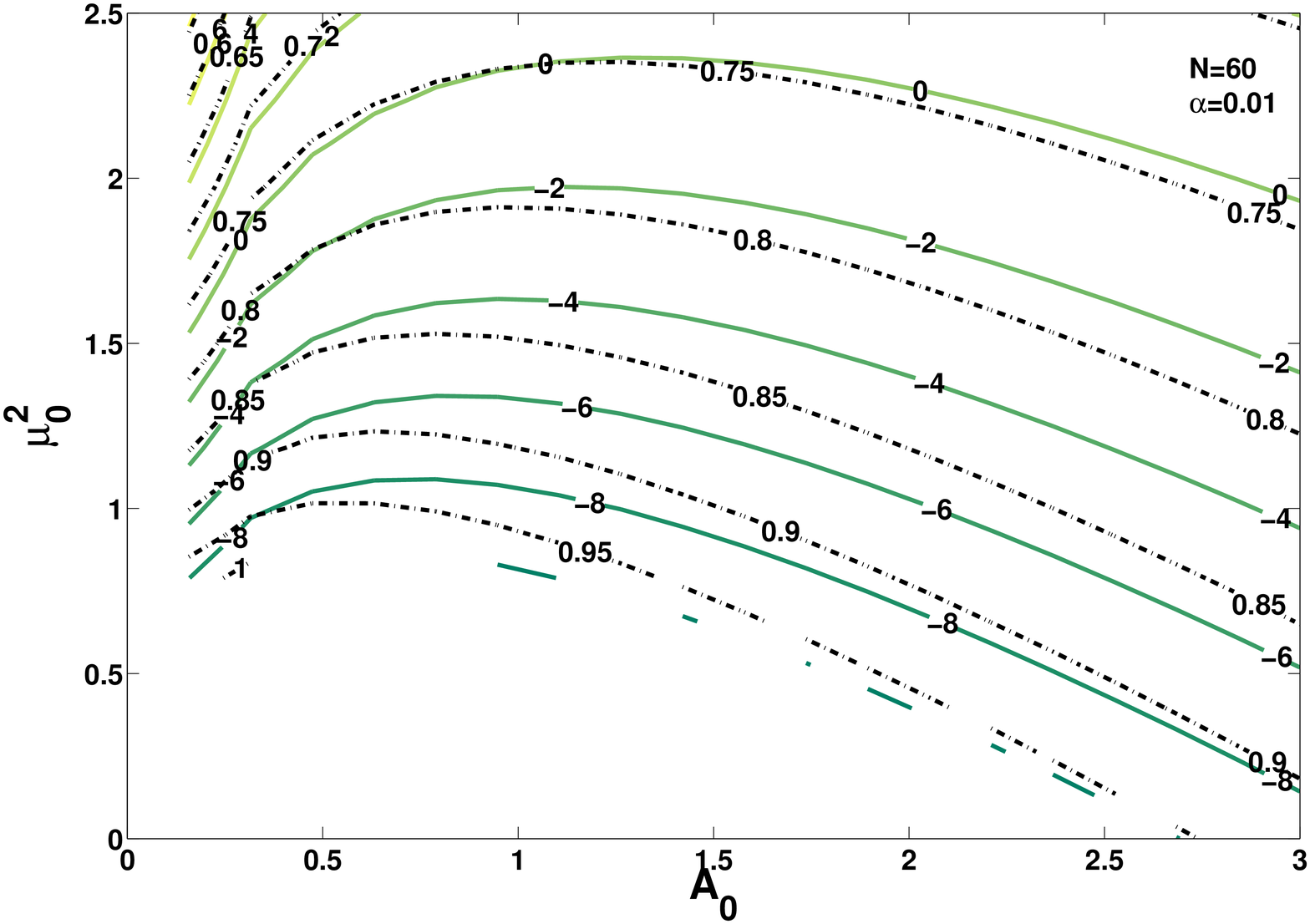}
\end{minipage}
\caption{In these plots we fix $N$ and $\alpha$, plotting contour lines 
of the spectral index (dashed) and $\log(\mathcal{B})$ (solid). Note that the contour lines
do not exactly match Ref.~\cite{Leach:2000ea}, an anomaly that we discuss in the text. Parameter space below $\log(\mathcal{B})\sim-8$ is excluded.}
\label{leach}
\end{figure*}

From \fig{leach} we find that in order to avoid the
overproduction of primordial black holes, $n>1$ would be ruled out
for $N=45$, and this bound is strengthened to ruling out 
$n>0.95$  for $N=60$. These bounds are slightly stronger than those
found by \cite{Leach:2000ea} who rule out $n\gtrsim1.1$ for $N=45$. 

However, there is a discrepancy between our spectral
index contour lines and \cite{Leach:2000ea}, which we found
was resolved by evolving our system an extra $\sim7$
$e-$folds. Via a process of elimination, we think this 
may be due to the fact that Ref.~\cite{Leach:2000ea} solved the background
equations numerically without resorting to the slow
roll approximation, while using the extended slow roll
formalism of \cite{Stewart:1993bc} to solve for the perturbations.
 
Either way, as our method underestimates 
the allowed parameter range for
each $n$, then using ref.~\cite{Leach:2000ea} method would strengthen our
conclusions i.e.our bounds are conservative.

\section{Discussion and Conclusions}\label{sec_discuss}

In this paper we utilised the spectrum on scales $\theta\lesssim0.3^o$,
corresponding to the end of inflation, to further the field of inflation model discrimination. 
The spectrum on these scales has yet to measured, but future
CMB surveys such as the PLANCK mission may constrain its' value. Astrophysical 
phenomena determines the upper bound to be $\calp_{\zeta_e}\sim10^{-4}$, 
corresponding to the criteria for the formation of primordial black holes (PBHs). 
In terms of the $\epsilon$ slow roll parameter, this means that the value 
of $\eps$ at the end of inflation must be much smaller than it's value at horizon
exit (\ref{pbh_cond}). The models of inflation that satisfy this condition must 
therefore exhibit the unique property of having a flatter slope towards the end of inflation.
So far, we only know of generic Hilltop models of inflation that fulfil this criteria.

We have investigated whether these generic models of Hilltop
inflation would lead to the production of primordial black holes with
a spectrum $\calp_{\zeta_e}\sim10^{-4}$.  We found that within the
range of parameters allowed by the latest WMAP data, the Hilltop model
(\ref{potential}) would lead to the formation of PBHs without
violating astrophysical bounds for $p<2.5$ and $q\leq3$ if $N>60$, and
for $p\sim2$, $2<q\ll3$ if $N>40$.  
Integral values of $p$ and $q$, which have some theoretical motivation,
only lead to PBH formation within the bound (\ref{pbh_cond}) for
$p=2$ and $q=3$, with $60\ll{}N<100$. In all cases it seems that near maximal running
is required. If, however we were to allow $N\gg110$ the
range of parameters that would lead to PBH formation would be extended.

\begin{figure*}
 \centering
\includegraphics[totalheight=2.5in,width=\linewidth]{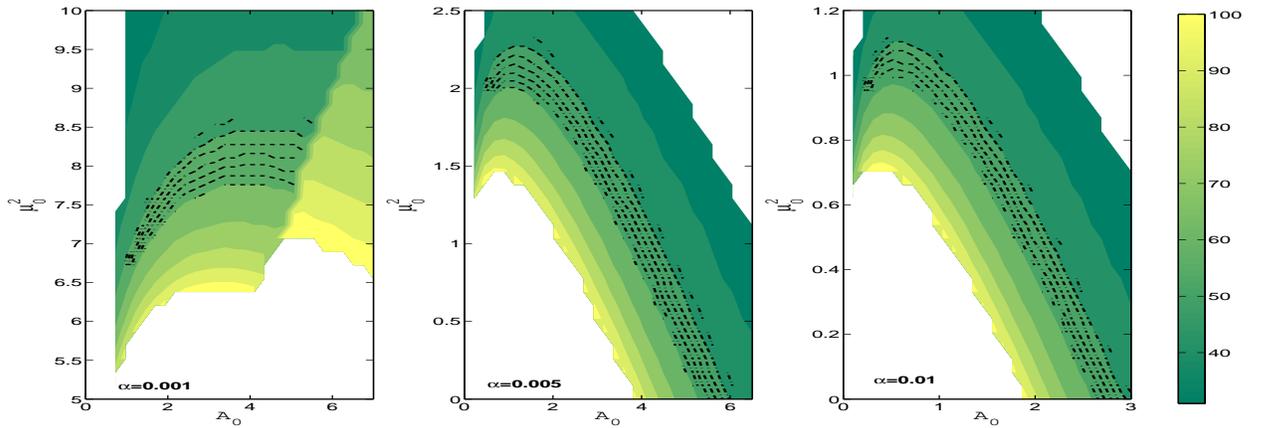}
\caption{Contour plots of the number of $e-$folds produced in the running mass model for three values of the gauge coupling
 $\alpha=[0.001,0.005,0.01]$. We found that $\alpha=0.1$ did not satisfy the WMAP bounds on $n_s$ and $n'$. We have filtered out the allowed parameter space for $10\leq{}N\leq110$, and coloured it in shades of green. The dashed regions in each plot correspond to the more `reasonable' bound $N=54\pm7$}
\label{Nreasonable2_RMM}
\end{figure*}

The allowed parameter range for the production
of primordial black holes with $\calp_{\zeta_e}\lesssim10^{-4}$ in the running mass model
is again dependent on $N$ as can be seen from \fig{Nreasonable2_RMM}.
We find that for $\alpha=0.01$ and $\mu_0^2\gtrsim1.1$, black holes \emph{could} form
after $N<47$ $e-$folds, and therefore before what
can be considered a `reasonable' end to inflation. This is problematic
on two counts, if we assume that the PBHs formed prior to the end of
inflation, then this could lead to the overclosure of the universe (c.f. \cite{Khlopov:2004tn}). 
On the same note, we know that on CMB scales
the spectrum is too small to support PBH production. On the second count,
assuming that the formation of the PBHs coincided with the end of inflation,
then the arguments we presented in section (\ref{N_sec}) apply. Thus,
using $N$ as a discriminator we rule out $A_0>3$ and $\mu_0^2>1.1$ for $\alpha=0.01$,
$A_0>6$ and $\mu_0^2>2.4$ for $\alpha=0.005$, and $A_0>5$ and $\mu_0^2>8.75$ for $\alpha=0.001$.
As we mentioned in the text $\alpha=0.1$ is ruled out on WMAP consistency grounds.

This model has also been analysed by \cite{Bugaev:2008gw}, in which the
authors use neutrino and $\gamma-$ ray background data to constrain the PBH mass spectrum,
which determines the spectral index on small scales $k\sim{}15~\rm{Mpc}^{-1}$. 
Then assuming that the running mass model is correct they reconstruct the power spectrum, 
finding that on these small scales the spectrum is highly sensitive to the running of the spectral index. 
Combining these two pieces of information they get bounds on $n_s$ and $n_s'$, 
which turn out to be inclusive of the WMAP limits on these parameters. That is, by 
using a different approach to ours \cite{Bugaev:2008gw} conclude that the 
running mass model is consistent with WMAP while avoiding PBH over-production, 
congruously with our findings.

Finally, we note that our generic results are consistent with the
findings of \cite{Peiris:2008be}. They tackle the question
of PBH production utilising the arising constraints to derive
bounds on the cosmological parameters, and
conclude that the PBH constraint is `strongly' dependent on $N$ and
the spectrum at the end of inflation. Characteristics exhibited by our specific models.

\section{Acknowledgements}
We thank Andrew Liddle, James Lidsey, David Lyth, Karim Malik, Hiranya Peiris, and David Seery
for useful comments and discussion. LA is supported by the 
Science and Technologies Facilities Council (STFC) under Grant 
PP/E001440/1. K.K. is supported in part by STFC grant,
PP/D000394/1, EU grant MRTN-CT-2006-035863, and the European Union through
the Marie Curie Research and Training Network ``UniverseNet''.

\bibliographystyle{apsrev}
\bibliography{pbh_hilltop}

\end{document}